\documentstyle[12pt]{article}

\newcommand\beq{\begin{equation}}

\newcommand\eeq{\end{equation}}

\newcommand\parm{\par\medskip}

\newcommand\ul{\underline}

\newcommand\be{\begin{eqnarray}}
\newcommand\ee{\end{eqnarray}}
\newcommand\gim{i8q/\gamma M}

\begin{document}

\centerline{\bf Scattering in Soliton Models and Boson Exchange Descriptions}
\vskip 0.6in
\centerline{C.\, Corian\`{o} , R.\, Parwani, H.\, Yamagishi ,  I.\, Zahed }
\vskip 0.4in
\centerline{ Department of Physics}
\centerline{State University of New York at Stony Brook}
\centerline{Stony Brook, New York 11794, USA}
\parm

\vskip 1.5 in
\centerline{Phys. Rev. D, in press}
\vskip .5cm
\centerline{\bf ABSTRACT}
\vskip .5cm
We argue that the description of meson-nucleon
dynamics based on the boson-exchange approach, is compatible
with the description of the nucleon as a soliton in the
nonrelativistic limit. Our arguments are based on an analysis of
the meson-soliton form factor and the exact meson-soliton
and soliton-soliton scattering amplitudes in the Sine-Gordon model.

\par\vfill

\centerline{Supported by the Department of Energy
under Grant No.\ DE-FG02-88ER40388}
\centerline{with the State University of New York.}
\newpage

{\bf 1. Introduction}
\vskip .4cm

The idea that the nucleon is a chiral soliton has attracted considerable
attention in the past few years [1]. This idea is supported by large $N_c$
counting arguments in QCD [2]. While the exact effective meson dynamics of
which the
nucleon is a soliton is not known, the generic character and properties
of the low-lying baryons can be understood from simple chiral models of
the Skyrme type.

The underlying degrees of freedom in the Skyrme model
are pions subject to the tenents of chiral symmetry. The model is consistent
with current algebra, and produces solitons that resemble ordinary nucleons.
The model provides a natural setting for analysing systematically
dynamical issues related to meson-nucleon and nucleon-nucleon scattering.
This is a distinct advantage over relativistic bag models and nonrelativistic
constituent quark models.

The meson-nucleon scattering amplitude can be systematically expanded
in powers of $1/\sqrt{N_c}$ around the classical one soliton solution.
The order $(1/\sqrt{N_c})^0$ has been
investigated by many authors and yields phase shifts that are overall
consistent with the data in the higher partial waves [3]. The expected
shortcomings of the P-wave amplitudes have been overcome at
next-to-leading order [4]. The deficiencies in the S-wave amplitudes are
unavoidable [5].

The nucleon-nucleon scattering amplitude has been more of a challenge
since the classical two soliton solution is not known. To leading order
in $1/\sqrt{N_c}$ a product ansatz has been used to construct the
classical two skyrmion potential. The scattering amplitude to leading order
follows automatically. The results of this approach compare favorably with the
empirically motivated potentials at short and long distances, but fail to
reproduce the intermediate range attraction in the scalar channel [6].
The numerical attempts to the scattering problem involve asymptotic
skyrmions and do not seem to relate immediately to the nucleon-nucleon
problem [7].

Recently two of us [8] have argued that the nucleon-nucleon
problem in the context of Skyrme models can be systematically analysed
using a double expansion in both the range $e^{-m_\pi r}$ and
the coupling constant $1/\sqrt{N_c}$. As a result the ambiguities related
to the ansatz dependence are lifted through the pion fluctuations,
and attraction is seen in the central potential at the two pion range.
Undoubtedly, this is a step forward in the process of describing unambiguously
the two-nucleon problem in the context of soliton models.

The purpose of the preceding discussion was to emphasize the consistent
framework offered by realistic soliton models in the description of a variety
of hadronic phenomena. It also provides a systematic framework for calculations
based on the semiclassical expansion.

To what extent
this description is compatible with the conventional boson-exchange
approach remains still unclear. In fact, simple arguments based
on naive power counting seems to run into  difficulties [8]. It is
the purpose of this paper to try to clarify some of these issues in the
context of completly integrable models. For definiteness we will use
the Sine-Gordon model. In this model many exact non perturbative results
are known [9,10], in particular an exact S-matrix [11]
 and all the form factors [12] are available.

The paper is organised
as follows : in section 2, we will analyse the meson-nucleon form factor
to one-loop order using the semiclassical expansion .The result
is shown to agree with the  exact result [12] in the
weak coupling (nonrelativistic) limit in appendix A.
In section 3, we discuss the nonrelativistic reduction of the exact
meson-soliton scattering amplitude.
In section 4, we discuss the exact soliton-soliton scattering amplitude
derived by Zamolodchikov and Zamolodchikov. We work out explicitly
the pole of the scattering amplitude in both the relativistic and
nonrelativistic limit. In section 5, we show how one can model the
nonrelativistic scattering amplitudes using simple boson-exchange models.
Our conclusions will be summarized in section 6.

\vskip 1.cm
{\bf 2. Meson-Soliton Form Factor}
\vskip .4cm

Consider the Sine-Gordon Lagrangian

\be
L = \int dx \left( \frac 12 (\partial_\mu\phi)^2 +
     \frac{m_0^2}{g^2} ({\rm cos}(g\phi ) -1 )\right)
\ee
with the classical one soliton solution

\be
\phi_s (x) =\frac 4g{\rm tan}^{-1}\left( e^{+m_0x}\right)
\ee
where $g$ is a small parameter ($g^2 <<8\pi$).
The bare soliton mass is $M_s=8m_0/g^2$,
and to leading order, the semiclassical description is justified.
Besides the soliton, there are single meson states
and breather modes  with masses in the
range $g^0\rightarrow g^{-2}$

\be
m_n =2M_s \,{\rm sin} (\frac {n}{16}\,g^2 )\qquad\qquad
                                n=1,2,... < \frac{8\pi}{g^2}
\ee

As a first step towards understanding meson-soliton dynamics we will discuss
the structure of the meson-soliton form factor

\be
F(Q) = <p_2 |\phi (0) |p_1>
\ee
where $|p>$ refers to a momentum eigenstate of the soliton,
with $Q=p_1-p_2$. To leading order, the form factor is of order $g^{-1}$.
To evaluate (4)
we will use the collective coordinate method [13], although other methods are
possible. For that, consider

\be
\phi (x,t) = \phi_s (x - X(t)) + \xi (x-X(t),t)
\ee
where $X(t)$ will be treated as a collective coordinate conjugate
to the total momentum of the soliton. In the semiclassical limit,
the meson fluctuations in (5) can be ignored and the result for
the form factor to leading order reads (Breit-frame : $Q = (0,q)$)

\be
F(q) = \int_{-\infty}^{+\infty}\, e^{iqx}\,\phi_s(-x)\,dx
     =-\frac{2i\pi}g \frac 1q \, {\rm sech}(q\pi/2m_0)
\ee

In the time-like region (6) displays
a string of odd poles located at $q_n=im_0(2n+1)$ with $n=0,1, ...$.
The pole at $q =0$ is of kinematical origin. It accounts for the
topological charge. Indeed, if we recall that the topological (baryon)
current is given by $J^\mu=\epsilon^{\mu\nu}\partial_\nu\phi$, then  $iqF(q)$
accounts for the proper charge  at $q=0$ (here unity). This point is suggestive
of a derivative meson-soliton coupling as given by (8) below.  The odd poles
correspond to the one-meson and breather modes
with the expected masses (6). The occurence of only
odd poles in (3) suggest that the charge conjugation of the
original meson field is odd.

The meson-soliton Yukawa coupling $f_0$  is defined by
\be
f_0 (q^2)= (q^2+ m_0^2)\, iqF(q)
\ee
Its value at the one meson pole is $f_0 (-m_0^2)=-8/g $. This
suggests that the meson-soliton coupling is strong and of pseudovector
nature. For point like solitons, we have

\be
{\cal L} = -i\frac 8g\overline\psi\gamma^\mu\gamma_5\psi\,\partial_\mu\phi
\ee
where $\psi$ is a fermion field of mass $M_s$. Note that in this form,
the pseudovector coupling is strong.

To estimate the role of the quantum effects on the structure of the
 meson-soliton
form factor we will evaluate the one-loop correction to (6).  For that, we will
use the collective quantization method discussed  by Tomboulis
[13] (and references therein). For the Sine-Gordon model,
the renormalised Hamiltonian  to order $g^2$ reads

\be
H =M_s+H_0 +H_1 +{\cal O} (g^2)
\ee
where

\be
H_0 =  \int dx \, \left( \frac {\pi^2}2 +\frac {{\xi '}^2}2
          +\frac {m_0^2\xi^2}2 (2\,{\rm tanh}^2(m_0x) -1)\right)  \nonumber \\
H_1 =  \frac 13\, gm_0^2\,\int dx \, \xi
               \left( \xi^2-\frac 3{4\pi}\int\frac{dk}{\sqrt{k^2+1}}\right)
                     \frac{{\rm sinh}(m_0x)}{{\rm cosh}^2(m_0x)}
\ee
In $H_1$ the term linear in the meson coupling is the mass counterterm
following normal ordering in the meson sector. To this order, the soliton
does not recoil. The meson field is quantized in a box of length L,

\be \xi (x,t)=\sum_{n}{1\over {2\omega_n}}\left( b_n\psi_n(x)e^{-i\omega_n t}
+b^{+}_n{\psi_n}^{+}(x)e^{i\omega_n t}\right)\ee
with ($\omega^2_n = m_0^2+(m_0q_n)^2 $)

\be
\psi_n(x)={{({\rm tanh}(m_0 x)-iq_n)e^{im_0 q_n x}}\over
{[(1+q_n)^2 L -{2/ m_0}{\rm tanh}({m_0L/2})]^{1/2}}}
\ee

The one-loop correction to the semiclassical expression (8) for
the form factor can be obtained using time-independent Rayleigh-Schrodinger
perturbation theory. Generically

\be
<p_2 |\phi (x) |p_1>_g =
\sum_m
\left({{ <p_2|H_1|m><m|\phi(x)|p_1>}\over{E_0 -E_m}}
+ {{<p_2|\phi(x)|m><m|H_1|p_1>}\over{E_0 -E_m}}\right) \nonumber \\
{\,\,\,}
\ee
Using the decomposition (5) for the full quantum field together
with the quantized meson fluctuations (11-12), we obtain

\be
<p_2 |\phi (x) |p_1>_g  ={\lim_{L\to\infty}}
   \int\, dz\,dy\,e^{iqz}{\rm sin}(g\phi_s (y))\,A_L(x,y,z)B_L(y)
\ee
where the integrands are given by

\be
A_L=&&\sum_n\frac 1{2\omega_n^2}
    \left( \psi_n (x-z)\psi_n^+(y) +\psi_n(y)\psi_n^+(x-z) \right)\\
B_L=&&\frac {m_0^2g}4\left(
                     \sum_n \frac 1\omega_n\psi_n(y)\psi_n^+(y)
                    -\frac 1{2\pi}\int\frac{dk}{\sqrt{k^2+1}}\right)
\ee
and $\phi_s$ is the soliton profile (2).
In the continuum limit (${\rm L}\rightarrow\infty$) the discrete sums can be
turned into continuous integrals with the proper weight for the meson
density of states around a non-moving soliton
(to the order quoted the soliton does not recoil), $i.e.$

\be
\rho (k) = \frac{m_0{\rm L}}{2\pi}-\frac 1{2\pi}\delta '(k)
\ee
where $\delta(q_n)=-2{\rm arctan}(q_n)$, denotes the phase shift of the box
eigenfunctions (12). As a result

\be
B_\infty = \frac{m_0^2g}{8\pi}\int\frac {dk}{(k^2+1)^{3/2}}
           ({\rm tanh}^2(m_0y) -1 )
\ee
Note that the result is finite (as it should) following the cancellation
between the divergence in the mode sum and the mass counterterm in (10).
This provides an additional check that mass renormalization
in the trivial vacuum sector is sufficient to renormalise the model
in the non trivial topological sector as well.
Similarly,

\be
A_\infty = \frac 1{4\pi m_0}\int dk\,\frac{e^{im_0k (x-y-z)}}{(k^2+1)^2}
    (k+i{\rm tanh}(m_0 (x-z) ))(k-i{\rm tanh}(m_0y) )
     +\,\,\,{\rm c.c.}\,\,\,
\ee
Inserting (18-19) in (14) and performing some algebra, we obtain

\be <p_2|\phi(x)|p_1>_g ={{m_0 g}\over {4\pi^2}}
\int_{-\infty}^{\infty}{\rm dz} e^{iqz}
\int_{-\infty}^{\infty}
{{\rm dk}\over {(k^2 +1)^2}}\int_{-\infty}^{\infty}dy
{\rm{sinh}(m_0y)\over{\rm{cosh}^4(m_0y)}} G_1(x,z,k,y)  \nonumber\\
{\,\,\,}
\ee
where
\be G_1(x,z,k,y)=\left[ {\rm th}(m_0(x-z))th(m_0y) + k^2\right]
{\rm cos}(m_0 k(x-z-y))\nonumber \\
-k\left[{\rm th}(m_0(x-z))-{\rm th}(m_0y)\right]{\rm sin}(m_0k(x-z-y))
\ee
The integrals are performed in the order indicated, so that no spurious
divergence is generated. Using the result

\be \int_{0}^{\infty}{\rm dt}\frac{{\rm cos}(at)} {{\rm ch}^{\nu}(bt)}
=\frac{2^{\nu -2}}{b\Gamma(\nu)}
\Gamma\left({\nu\over 2}+{ai\over 2b}\right)
\Gamma\left({\nu\over 2}-{ai\over 2b}\right) \ee
and integrating by parts, gives

\be
<p_2|\phi (x) |p_1>_g =  (\frac{-ig}{8q})e^{iqx}{\rm sech}(\frac{q\pi}{2m_0})
(\frac{q^2}{{m_0}^2})
\ee
Combining (6) with (23) yields the form factor to order $g^2$

\be
<p_2 |\phi (0) |p_1> =-\frac {i2\pi}{q}{\rm sech}(\frac {q\pi}{2m_0})
\left(\frac 1g +\frac{gq^2}{16\pi m_0^2}+{\cal O}(g^2)\right)
\ee
While the pole position has not changed, the residue has been modified.
The quantum effects are expected to renormalize the bare position of the
pole (here meson mass) and affect the strength of the residues.
In fact we expect this result to hold true
to higher order in perturbation theory
around the soliton background and even in higher dimensions.

\vskip 1.cm
{\bf 3. Meson-Soliton Scattering}
\vskip .4cm

The exact meson-soliton scattering amplitude
is given by [9],

\be
S_1 (\theta_-) =\frac {{\rm sinh}\theta_- +i{\rm cos}(\gamma/16)}
                      {{\rm sinh}\theta_- -i{\rm cos}(\gamma/16)}
\ee
In the meson-soliton center of mass frame $p+k =0$, the scattering amplitude
(25) can be rewritten  in momentum space

\be
S_1 (k) = \frac {k (\sqrt{m^2+k^2}-\sqrt{M^2+k^2}) -imM{\rm cos}(\gamma/16)}
                {k (\sqrt{m^2+k^2}-\sqrt{M^2+k^2}) +imM{\rm cos}(\gamma/16)}
\ee
In the semiclassical (nonrelativistic) limit $\gamma\sim g^2$, $M\sim M_s$,
$m\sim m_0$  and (26) reduces to

\be
S_1 = 1 +\frac{2im_0k-2m_0^2}{k^2+m_0^2} +{\cal O}(g^2)
\ee

On the other hand, the scattering process of a meson off a soliton in the
semiclassical description corresponds to background field scattering off
a static soliton. If $\psi_k (x)$ designates the continuum scattering
wavefunction of the meson, then from (12) we have

\be
\psi_k (x) = \frac 1{\sqrt{2\pi}}\frac k{\omega_k} e^{ikx}
\left( 1+\frac {im_0}k {\rm tanh}\,(m_0x) \right)
\ee
where $\omega_k =\sqrt{m_0^2+k^2}$,  $-\infty<k<\infty\,\,\,.$
The S-matrix to order $g^2$ follows from (28) through the identification

\be
S_p = \frac {\psi_k(+\infty ) }{\psi_k(-\infty )}
\ee
A little algebra shows that this agrees with (27). Here, we point out that
with the Yukawa coupling defined as in (7), it was shown in [14-15]
that the Born diagrams for meson-soliton
scattering give the same result as potential scattering (29).
Our result, while in agreement with [14-15], shows that (29) follows
directly from  the exact scattering amplitude in the semiclassical
limit, as it should.

\vskip 1.cm
{\bf 4. Soliton-Soliton Scattering}
\vskip .4cm

The exact relativistic S-matrix for the Sine-Gordon model has been
derived by Zamolodchikov  and Zamolodchikov [11]. The form of the S-matrix
follows from factorisation (absence of pair creation), unitarity and
crossing. The soliton-soliton S-matrix element reads
\footnote{We are using plane-wave normalisations}

\be
{}_{+}<p_1'p_2' |p_1p_2>_{-} =  {}_{-}<p_1'p_2' |p_1p_2>_{-} \,S(\theta_-)
\ee
where $S$ depends only on the relative rapidities $\theta =\theta_-$

\be
S(\theta) ={\rm exp}\left(i\pi -i\int_0^\infty\,dk\,
\frac{{\rm sin}(k\theta )}{k}
\frac{{\rm sinh}((\pi-\zeta)k/2)}{{\rm cosh} (\pi k/2){\sinh} (\zeta k/2)}
\right)
\ee
with $\zeta =\gamma/8= g^2/(8-g^2/ \pi)$ and
$\theta_{\pm} =\theta_1\pm\theta_2$.
The rapidity variables $\theta_{1,2}$ are related to the momentum variables
$p_{1,2}$ as follows

\be
p_{1,2} =M ({\rm cosh}\theta_{1,2} , {\rm sinh}\theta_{1,2})
\ee
Here $M=8m/\gamma$ is the renormalized soliton mass and $m$ the renormalized
meson mass.

Alternative representations of (30-31) can be found in Ref. [10]. In the
s-channel, the $S$-matrix exhibits a treshold (branch point) at $s=4M^2$ and
bound states (poles) for $0<s<4M^2$. As a result, the soliton-soliton amplitude
is a meromorphic function of $\theta$ in the strip $0<{\rm Im}\theta <\pi$
following the mapping $s=4M^2{\rm cosh}^2(\theta /2)$. The edges of the strip
correspond to the cuts in the $s$-plane. Crossing symmetry follows from
$\theta\rightarrow i\pi -\theta$.

To exhibit the singularity structure of (31), it is best to define
$\Omega (\theta)$, a regular function in the strip
$0<{\rm Im}\theta <2\pi$ that satisfies

\be
S(-\theta )=\Omega (\theta )\,S(\theta ) =\Omega (\theta -i2\pi )\\
\Omega (\theta )\Omega (\theta -i\pi )= \eta^{-1}(\theta +i\pi/2 )
\ee
where $\eta (\theta )$ is an even function with simple poles
at $\theta_n=i\pi/2+in\zeta $ with $n\geq 0$ and zeroes at
$\theta_n =i3\pi/2 +in\zeta$ with $n\geq 1$. For more details
we refer to Smirnov [12]. The auxiliary function $\eta (\theta )$
satisfies the property

\be
\eta (\theta -i\zeta ) =\eta (\theta )
\frac{{\rm cosh}((\theta +i\pi /2)/2)}
     {{\rm cosh}((\theta -i\pi /2-i\zeta )/2)}
\ee
In terms of (33-34) we can rewrite the scattering amplitude (31)
in the following form

\be
S(\theta ) =\frac{\eta (\theta +i\pi /2)}{\eta (\theta -i\pi /2)}
\ee
Using (35) we obtain

\be
S(\theta ) =S(\theta -i\zeta ){\rm coth}(\theta /2)
                              {\rm coth}((\theta -i\zeta )/2)
\ee
 Iterating (37)we easily find ($n\geq 2$)

\be
S(\theta) =S(\theta-in\zeta ){\rm coth}(\theta /2) \,{\rm H}_n^2(\theta )\,
                           {\rm coth}((\theta -in\zeta )/2)
\ee
where we have defined

\be
{\rm H}_n(\theta ) = \prod_{k=1}^{n-1}
                     {\rm coth}((\theta-i(n-k)\zeta )/2)
\ee
The expression (38) for the scattering amplitude displays a string of
simple poles at $\theta_n = in\zeta=in\gamma /8$ with residues given by

\be
R_n = -2{\coth}{(\theta_n /2)}\,{\rm H}_n^2 (\theta_n )
\ee

To extract the semiclassical limit it suffices to notice that this limit
is the same as the nonrelativistic limit for which the rapidity becomes
the ordinary velocity, $\theta\rightarrow v=p/M_s$. With this in mind
the $S$-matrix in the semiclassical limit (Breit-frame), near the
$\theta=\theta_n$ pole
reads

\be
S= \frac {M_sR_n}{q-q_n}
\ee
with  the location of the poles and residues given by

\be
q_n=inm_0\qquad\qquad M_sR_n=(-1)^{n+1}\frac {in}{(n!)^2}(\frac{16}{g^2})^{2n}
\,\,m_0
\ee
{\,\,\,}
The nonrelativistic soliton-soliton S-matrix in the Sine-Gordon model
can also be obtained by solving the factorisation equation and demanding
unitarity. The result is [11] (in momentum space)
\be
S(q)={{{\rm sh}(i\pi\alpha -{{8\pi q}/ {\gamma M}})
 \Gamma(-\gim -\alpha)\Gamma(-\gim +\alpha +1)}\over
 {{\rm sh}({{\pi q}/ m})\Gamma(-\gim)\Gamma(1-\gim)}}
\ee
The constant $\alpha$ is not fixed in (43) because crossing symmetry
is not required in the nonrelativistic case.
We fix it here by demanding that in the weak coupling limit
(i.e.$ \gamma\sim g^2<<1$ )the poles and residues of (43)
match (41-42). A simple calculation gives
\be \alpha=\frac{16}{g^2}
\ee
The amplitude (43) implies a potential between the solitons
of the form
\be
V(r)=\frac{M\gamma^2}{64}{{\alpha^2 -\alpha +3/4}\over
                 {{\rm sinh}^2(({m\gamma}/{16})r)}}
\ee
In the weak coupling limit ($g^2 <<1$), we get

\be
V(r)=\frac{32m_0}{g^2}{1\over{{\rm sh}^2{{m_0 r}/ 2}}}
    =128m_0 \frac{e^{-m_0r}}{g^2} \sum_{n=0}^{\infty}e^{-nm_0 r}(1+n)
\ee
which is suggestive of a one-boson exchange description
of the soliton-soliton scattering in the low-momentum regime.
This was also indicated by the pole structure of the S-matrix (41-42).
In the next section we will investigate how well the postulated
effective interaction (8) supports this picture.
\vskip 1.cm
{\bf 5. Boson-Exchange Models}
\vskip .4cm

{}From the meson-soliton form factor analysis of section 2, we have concluded
that in the nonrelativistic limit, meson-soliton dynamics suggests
using a pseudovector coupling of the form given by (8).


 The soliton-soliton scattering amplitude follows from
the direct and exchange diagrams shown in Fig. 1,

\be
<p_1'p_2' |(S-1)|p_1p_2> = (2\pi)^2
\frac{\delta^2(p_1'+p_2'-p_1-p_2)}{\sqrt{2E_12E_22E_1'2E_2'}} \,T (s,t,u)\,
\ee
where $T$ is the pertinent  T-matrix amplitude.
The restricted character of the kinematics
in $1+1$ dimensions  and the mass-shell conditions yield

\be
\delta^2(p_1'+p_2'-p_1-p_2) =
\frac{E_1E_2}{|p_1E_2-p_2E_1|}
\left( \delta (p_1'-p_1)\delta (p_2' -p_2) +
       \delta (p_1'-p_2)\delta (p_2' -p_1) \right)
\ee
Combining (47) with (48) and going to the center of mass frame
$p_1=-p_2=p$ ,$E_1=E_2=E=\sqrt{M_s^2+p^2}$ and $q=p_1-p_2=2p$ give

\be
<p_1'p_2' |(S-1)|p_1p_2> = &&(2\pi)^2
\left( \delta (p_1'-p_1)\delta (p_2' -p_2) -
       \delta (p_1'-p_2)\delta (p_2' -p_1) \right) \nonumber\\
&&(-\frac {64}{g^2} \frac {M_s^2}E)\,\frac q{q^2+m_0^2}
\ee
As expected there is a pole at $q=im_0$ (one-meson exchange) with
a residue equal to ($n=1$)

\be
M_sR_1 = i\frac {256}{g^4}\,\,m_0
\ee
This is in total agreement with the semiclassical result obtained
from the exact S-matrix. It is not hard to show that additional
Yukawa-couplings corresponding to the breather modes (bound meson states)
in the form of vector or pseudovector couplings
(depending on their intrinsic charge conjugation) with the appropriate
strength, reproduce the scattering amplitude (43). In fact the character
of the residues in (42) is suggestive of the following hierarchy of
couplings (generically)

\be
\frac 1g \phi \overline\psi\psi\qquad ,\qquad
\frac 1{g^2} \phi\phi \overline\psi\psi\qquad ,\qquad
\frac 1{g^3} \phi\phi\phi \overline\psi\psi\qquad ...
\ee
They are to be compared with the ones expected in chiral models where
$g\rightarrow f_\pi$ and $\phi\rightarrow \pi$ which are reminiscent
of pion-nucleon, rho-nucleon, omega-nucleon .... couplings.
We conclude, that in the semiclassical limit the conventional boson-exchange
approach is compatible with the soliton-soliton scattering description.

\vskip 1.cm
{\bf 6. Conclusions}
\vskip .4cm

We have analysed soliton-soliton and meson-soliton scattering amplitudes
in the semiclassical limit. The exact results derived by Zamolodchikov
and Zamolodchikov are in agreement with conventional descriptions
based on boson-exchange models in the semiclassical (nonrelativistic)
limit provided that the couplings are chosen appropriately. We have also
shown that the meson-soliton vertex, while in agreement with the exact
results of Smirnov in the semiclassical limit, provides important insights
on the meson-soliton dynamics to leading order. We expect the present
conclusions to carry through to higher space-time dimensions in more realistic
models such as the Skyrme model.


\newpage

{\bf  Appendix A : Smirnov's result in the weak coupling limit}
\vskip .4cm

The semiclassical description of the meson-soliton form factor and
the one-loop correction discussed in section 2, are  consistent with
the exact result in [11-12]
\be
F(Q) = -\frac{i2\pi}g \frac {e^{J(\theta_-)}}q
      \left(\frac{{\rm cosh}\theta_+/2}
                 {\sqrt{ {\rm cosh}\theta_1{\rm cosh}\theta_2}}\right)
       \frac{ {\rm cosh}(\theta_-/2) }{ {\rm cosh} (4\pi\theta_- /\gamma)}
\ee
where

\be
J(\theta_-) = \int_0^\infty\,\frac{dx}x\,
\frac{{\rm sin}^2(x\theta_-/2) {\rm sinh}(x(\pi-\zeta)/2)}
     {{\rm cosh}(\pi x/2){\rm sinh}(\pi x){\rm sinh}(3x/2)}
\ee
The bracket in (52) involves a relativistic kinematic factor (numerator)
and energy normalisation factors (denominator). The combination of the bracket
and $1/q$ is relativistically invariant. In the region
$0<{\rm Im}\theta_-<2\pi$, the function $J(\theta_-)$ is regular, so that the
exact poles of the form factor follow from the position of the poles in
${\rm cosh}^{-1}(4\pi\theta_-/\gamma )$, $i.e.$

\be
\theta_n =\frac {i\gamma}8 (2n+1) \qquad\qquad n=0,1,...
\ee
in agreement with the poles derived using the semiclassical approximation.

Using the substitution $\theta\rightarrow v=p/M_s$
in the semiclassical limit, we have

\be
J(\theta_-) =&& \frac {g^2}{16\pi}(\frac {q^2}{m_0^2})+{\cal O}(g^4)\\
{\rm cosh}(\frac {\pi\theta_-}{2\zeta})=&&
{\rm cosh}(\frac {\pi q}{2m_0}) +{\cal O}(g^4)
\ee
Inserting (55-56) into (52-53) we obtain after little algebra

\be
<p_2 |\phi (0) |p_1> =-\frac {i2\pi}{q}{\rm sech}(\frac {q\pi}{2m_0})
\left(\frac 1g +\frac{gq^2}{16\pi m_0^2}+{\cal O}(g^2)\right)
\ee
in agreement with the result (24) derived in section ${\bf 2}$
 using perturbation theory.
This result shows agreement between perturbation theory in the presence
of a single soliton background and the exact result to one loop
 level and extends the lowest order check first done in [9].

\newpage

\parindent 0 pt
{\bf References}
\par\bigskip

1. For reviews and further references,
   see
   I. Zahed and G. E. Brown, Phys. Rep. $\ul{142}$, 1 (1986);
   G. Holzwarth and B. Schwesinger, Rep. Prog. Phys. $\ul{49}$, 825 (1986);
   {\sl Chiral Solitons}, edited by Keh-Fei Liu (World Scientific,
   Singapore, 1987); {\sl Skyrmions and Anomalies},
   edited by M. Jezabek and M. Praszalowicz (World Scientific,
   Singapore, 1987).
\parm
2. G. `t Hooft, Nucl. Phys. $\ul{B72}$, 461 (1974);
   E. Witten, Nucl. Phys. $\ul{B160}$, 57 (1979).
\parm
3. B. Schwesinger, H. Weigel, G. Holzwarth and A. Hayashi,
   Phys. Rep. $\ul{173}$, 173 (1989), and references therein.
\parm
4. H. Verschelde, University of Gent preprint   (1990)
\parm
5. H. Yamagishi and I. Zahed, {\sl Chiral Soliton versus Soft Pion Theorems
   in large} $N_c$ {\sl QCD}, Stony Brook prepint 1989.
\parm
6. A. Jackson, A.D. Jackson and V. Paskier, Nucl. Phys. $\ul{A432}$, 567
(1985).
\parm
7. J.J.M. Verbaarschot, T.S. Walhout, J. Wambach and H. Wyld,
Nucl. Phys. $\ul{A461}$, 603 (1987).
\parm
8. H. Yamagishi and I. Zahed, {\sl Nucleon-Nucleon Hamiltonian
   from Skyrmions}, Phys. Rev. D in Press.
\parm
9. P. H. Weisz, Phys. Lett. $\ul{67 B}$, 179, (1977).
\parm
10. S. Coleman, Phys. Rev. $\ul{D 11}$  (1975) 2088.
\parm
11. A.B. Zamolodchikov and Al.B. Zamolodchikov, Annals of Physics
$\ul {120}$, 253 (1979).
\parm
12. F.A. Smirnov, in {\sl Form factors in Completely integrable
Models of Quantum Field Theory}, Singapore, World SAcientific, in press.
\parm
13. E. Tomboulis, Phys. Rev.$\ul{ D12}$ (1975) 1678
and references therein.
\parm
14. K. Kawarabayashi and K. Ohta, Phys. Lett.$\ul {216 B}$, 205 (1981).
\parm
15. Y. G. Liang, B. A. Li, K. F. Liu and R. K. Su, Phys. Lett. $\ul{243 B}$,
133 (1990)

\end{document}